\documentclass[aps,a4paper,showpacs,twocolumn,amsmath,amssymb,floatfix,prb,preprintnumbers,footinbib]{revtex4-1}
\usepackage{graphicx}
\usepackage{amsmath,amsfonts,xfrac}
\usepackage{color}
\usepackage{import}
\usepackage{hyperref}

\newcommand{\ddt}[1]{\frac{d{#1}}{dt}}
\newcommand{\mrm}[1]{\mathrm{#1}}
\newcommand{\mbf}[1]{\boldsymbol{#1}}
\newcommand{\mat}[1]{\mbf{#1}}
\newcommand{\limes}[2]{\lim\limits_{{#1}\rightarrow{#2}}}

\newcommand{\epsd}{\epsilon_\mrm{d}}
\newcommand{\epsdin}{\epsilon_\mrm{d}^\mrm{in}}
\newcommand{\epsSQD}{\epsilon_\mrm{SQD}}
\newcommand{\epsSQDin}{\epsilon_\mrm{SQD}^\mrm{in}}
\newcommand{\UC}{U}
\newcommand{\Ud}{U_\mrm{d}}
\newcommand{\GamL}{\Gamma_\mrm{L}}
\newcommand{\GamR}{\Gamma_\mrm{R}}
\newcommand{\GamC}{\Gamma_\mrm{C}}
\newcommand{\kBT}{k_\mrm{B}T}
\newcommand{\Vsd}{V_\mrm{sd}}

\newcommand{\POft}{\mbf{P}(t)}
\newcommand{\Pin}{\mbf{P}^\mrm{in}}
\newcommand{\Pst}{\mbf{P}^\mrm{st}}
\newcommand{\Pd}{\mbf{P}_\mrm{d}}

\newcommand{\fC}[1]{f^{#1}_\mrm{C}(\epsilon_\mrm{d})}
\newcommand{\fCUd}[1]{f^{#1}_\mrm{C}(\epsilon_\mrm{d}+\Ud)}

\newcommand{\falp}[1]{f^{#1}_\alpha(\epsilon_\mrm{SQD})}
\newcommand{\bfalp}[1]{\bar{f}^{#1}_\alpha(\epsilon_\mrm{SQD})}
\newcommand{\bbfalp}[1]{\bar{\bar{f}}^{#1}_\alpha(\epsilon_\mrm{SQD})}
\newcommand{\fR}[1]{f^{#1}_\mrm{R}}
\newcommand{\bfR}[1]{\bar{f}^{#1}_\mrm{R}}
\newcommand{\bbfR}[1]{\bar{\bar{f}}^{#1}_\mrm{R}}
\newcommand{\fL}[1]{f^{#1}_\mrm{L}}
\newcommand{\bfL}[1]{\bar{f}^{#1}_\mrm{L}}
\newcommand{\bbfL}[1]{\bar{\bar{f}}^{#1}_\mrm{L}}
\newcommand{\fd}[1]{f^{#1}_\mrm{d}}
\newcommand{\fdUd}[1]{f^{#1}_\mrm{d}(\Ud)}
\newcommand{\bfd}[1]{\bar{f}^{#1}_\mrm{d}}
\newcommand{\bfdUd}[1]{\bar{f}^{#1}_\mrm{d}(\Ud)}
\newcommand{\fS}[1]{f^{#1}_\mrm{SQD}}
\newcommand{\bfS}[1]{\bar{f}^{#1}_\mrm{SQD}}
\newcommand{\bbfS}[1]{\bar{\bar{f}}^{#1}_\mrm{SQD}}
\newcommand{\lamc}{\lambda_\mrm{c}}
\newcommand{\lamcp}{\lambda_+}
\newcommand{\lamcm}{\lambda_\mrm{\widetilde{c}}}

\newcommand{\lamsig}{\lambda_\sigma}
\newcommand{\lamsigp}{\lambda_\sigma^+}
\newcommand{\lamsigm}{\lambda_\sigma^-}
\newcommand{\lamsigpm}{\lambda_\sigma^\pm}
\newcommand{\lamSET}{\lambda_\mrm{\widetilde{SQD}}}
\newcommand{\blamc}{\bar{\lambda}_\mrm{c}}
\newcommand{\blamsig}{\bar{\lambda}_\sigma}
\newcommand{\lamm}{\lambda_\mrm{fp}}
\newcommand{\lamtm}{\lambda_\mrm{\widetilde{\mrm{fp}}}}

\newcommand{\DfS}{\delta\!f_\mrm{SQD}}

\newcommand{\Dfd}{\delta\!f_\mrm{d}}
\newcommand{\bDfS}{\delta\!\bar{f}_\mrm{SQD}}

\newcommand{\DfdUd}{\Dfd(\Ud)}

\begin{document}

\title{Detection of the relaxation rates of an interacting quantum dot by a capacitively coupled sensor dot }

\author{Jens Schulenborg$^{1,2}$, Janine Splettstoesser$^{1,2}$, Michele Governale$^{3}$, and L. Debora Contreras-Pulido$^{1,4}$}
\affiliation{$^{1}$Institut f\"ur Theorie der Statistischen Physik, RWTH
Aachen University, D-52056 Aachen, \& JARA - Future Information Technologies, Germany\\
$^{2}$Department of Microtechnology and Nanoscience (MC2), Chalmers University of Technology, SE-41298 G\"oteborg, Sweden\\
$^{3}$School of Physical and Chemical Sciences and MacDiarmid Institute for Advanced Materials and Nanotechnology, Victoria University of Wellington, PO Box 600, Wellington 6140, New Zealand\\
$^{4}$Institut f\"ur Theoretische Physik, Albert-Einstein Allee 11, Universit\"at Ulm, D-89069 Ulm, Germany
}

\date{\today}

\begin{abstract}
We present a theoretical study of the detection of the decay time scales for a single-level quantum dot by means of a capacitively coupled sensor dot, which acts as an electrometer. 
We investigate the measurement back-action on the quantum-dot decay rates and elucidate its mechanism. 
We explicitly show that the setup can be used to measure the bare quantum-dot relaxation rates by choosing gate pulses that minimize the back-action. Interestingly, we find that besides the charge relaxation rate, also the rate associated to the fermion parity in the dot can be accessed with this setup.
\end{abstract}

\maketitle

\section{Introduction}

 \begin{figure}[b]
\includegraphics[width=0.9\linewidth]{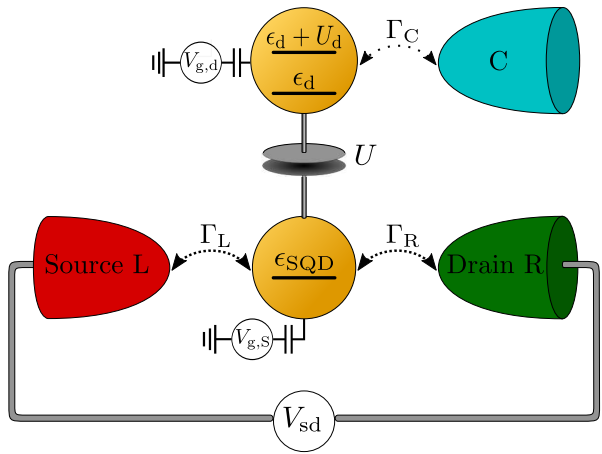}
\caption{
 \label{fig_simplemodel}
Sketch of the system. The upper dot (d) is driven out of equilibrium at time $t_0$; the lower dot (SQD) is used for detection of the relaxation process. The two single-level dots are coupled capacitively to each other and they are tunnel-coupled to different electronic reservoirs. Since the SQD electron spin is not important for the measurement, we neglect it for simplicity. }
 \end{figure}

Quantum dots driven by time-dependent signals play an important role for different nanoscale devices such as 
single-electron emitters,~\cite{Feve07,*Blumenthal07,*Kaestner08,*Hermelin11,*Mcneil11,*Pekola13} quantum pumps,~\cite{Kouwenhoven91,*Switkes99,*Buitelaar08,*Maisi09,*Giazotto11,*Roche13,Brouwer98,*Splett05,*Sela06,*Fioretto08} solid state qubits \cite{Loss98,Burkard99,Hanson03} and spin and charge read-out schemes.~\cite{Elzerman04} 
In order to assess the maximum frequency at which such devices can be operated, it is necessary to understand the relaxation behavior of quantum dots when they are driven out of equilibrium. Besides their relevance for nanoscale device operation, these types of  systems are  also interesting  from a basic-physics perspective, as they are a testbed to understand nonequilibrium and Coulomb-interaction
effects in size-quantized systems.~\cite{Reckermann10,Calvo12,Riwar13} Quantum dots are also used as one of the capacitor plates in a mesoscopic capacitor.~\cite{Buttiker93a} 
The theoretical analysis of charge relaxation in these devices, in the limit when the Coulomb interaction can be treated at the Hartree level, is based on the scattering approach to mesoscopic transport.~\cite{Buttiker93a,*Buttiker93b,*Buttiker96} In particular, the quantization of the  relaxation resistance has been extensively investigated both theoretically~\cite{Nigg06,*Nigg08,*Nigg09,*Mora10,*Ringel08} and experimentally.~\cite{Gabelli06} In recent works,~\cite{Rodinov09,*Hamamoto10,*Lee11,*Filippone11,*Filippone12,*Kashuba12} the theory has also been extended to \textit{strongly} interacting systems. This becomes especially important for quantum dots with only one relevant orbital level and strong Coulomb interaction. In such a single-level dot, the interaction furthermore leads to different decay time scales for the charge and the spin.~\cite{Splett10}  
Moreover, an additional decay rate has been predicted to govern the relaxation behavior of the system toward equilibrium.~\cite{Splett10,Contreras12} This rate does not contribute to the relaxation of the charge and spin, but can be linked to the fermion parity of the dot state. Interestingly, this fermion-parity rate depends neither on the level position nor on the interaction strength, and equals the sum of the bare tunneling rates for all transport channels even if higher-order tunneling processes are taken into account.~\cite{Contreras12,Saptsov12,Saptsov13}

The present paper is devoted to the problem of \textit{measuring} the different decay rates of a single-level quantum dot driven out of equilibrium by a step gate-voltage pulse. Particular emphasis will be given to the fermion-parity mode of the quantum dot, which has not been detected so far.  
Recent experiments were dedicated to the read-out of relaxation times of excited states~\cite{Fujisawa00,*Fujisawa01a,*Fujisawa01b,*Fujisawa02a,*Fujisawa02b,Wang10,Volk13} or the charge-relaxation rate of quantum dots.~\cite{Beckel14}
More generally, the (time-resolved) read-out of quantum-dot charge states has been investigated with charge sensors based on quantum point contacts,~\cite{Field93,Elzerman03,Ihn09,Kueng10,Neumann13} or based on a detector quantum dot (or metallic island).~\cite{Barthel10,Maisi11,Li07,Kiesslich11,Hell13}

In this paper, we study the read-out of relaxation rates using the setup shown schematically in Fig.~\ref{fig_simplemodel}: a single-level interacting quantum dot (d) attached to one electronic reservoir, brought out of equilibrium at a time $t_0$ by a step pulse of the gate voltage. For the detection of the relaxation behavior of this quantum dot, a second dot, acting as a sensor quantum dot (SQD), is coupled capacitively to the first one and tunnel-coupled to two further electronic reservoirs with a transport voltage applied between them.  Note that setups of this type have recently attracted interest in the context of feedback control~\cite{Strasberg13} and heat to current conversion.~\cite{Sanchez11} 
Here, the basic idea for the realization of a measurement is that the charge state of the quantum dot (d) will affect the current flowing through the SQD. However, the measurement set up will induce a back-action on the dot dynamics: the basic mechanism leading to the back-action is that the interdot Coulomb repulsion can induce Coulomb blockade in the quantum dot (d) and affect the relaxation rates. 
The main goals of this paper are: (1) understanding how the relaxation rates of the quantum dot (d) are modified by the measurement (back-action); (2) devising measurement protocols, which minimize the back-action effects and thus allow us to measure the \textit{bare} relaxation rates, that is, the quantum-dot relaxation rates in the absence of the SQD. 

The paper is structured in the following way. In Sec.~\ref{model}, we detail our model and review the formalism used to calculate the quantum-dot dynamics and the SQD current. 
Results are presented in Sec. \ref{results}. We start by discussing briefly the  bare relaxation rates of a single quantum dot in the absence of the SQD in Sec. \ref{sec_singledot} . We then consider the dynamics of the full d-SQD system and elucidate the measurement back-action effects in Secs. \ref{sec_dot-sensor} and \ref{sec_dynamics_measurement}. The measurement mechanism and the protocols to access the bare relaxation rates are examined in Sec. \ref{sec_measurement}. Finally, we summarize our findings in Sec. \ref{conclusions}.

\section{Model and Formalism}
\label{model}

\subsection{System Hamiltonian}\label{sec_hamiltonian}

The system under investigation is sketched in Fig.~\ref{fig_simplemodel}. It is described by the Hamiltonian
\begin{eqnarray}
H & = & H_\text{d}+H_\text{SQD}+H_\text{d-SQD}+H_{\text{leads}}+H_{\text{tun}},
\end{eqnarray}
where the terms on the right-hand side (r.h.s.) refer, respectively, to the dot, the SQD, the capacitive coupling between the two dots, the metallic leads, and the tunneling Hamiltonian, which describes the tunnel coupling between the dots and the leads. 
The Hamiltonians of the isolated two-dot system are
\begin{subequations}
\begin{align}
& H_\text{d} =  \sum_{\sigma=\uparrow,\downarrow}\epsilon_{\mrm{d},\sigma}d^\dagger_{\mrm{d},\sigma}d^{}_{\mrm{d},\sigma}+ \Ud n_{\mrm{d},\uparrow}n_{\mrm{d},\downarrow}\\
& H_\text{SQD}  =   \epsSQD d^\dagger_{\mrm{SQD}}d^{}_{\mrm{SQD}}\\
& H_\text{d-SQD} =   U n_\mrm{d}n_\mrm{SQD}\ ,
\end{align}
where $d^\dagger_{\mrm{d},\sigma}(d^{}_{\mrm{d},\sigma}$) is the creation (annihilation) operator for an electron with spin $\sigma$ on the dot and $ d^\dagger_{\mrm{SQD}}(d^{}_{\mrm{SQD}})$ the creation (annihilation) operator for an electron in the SQD. We also define the number operators as  $n_{\mrm{d}}=\sum_\sigma n_{\mrm{d},\sigma}=\sum_\sigma d^\dagger_{\mrm{d},\sigma}d^{}_{\mrm{d},\sigma}$ and  $n_{\mrm{SQD}}=d^\dagger_{\mrm{SQD}}d^{}_{\mrm{SQD}}$. 
The single-level energy of the dot (detector) is denoted by $\epsilon_{\mrm{d},\sigma}$ $(\epsSQD)$. The charging energy for a doubly occupied dot is $\Ud$, the energy $U$ denotes the capacitive coupling between the dot	 and the SQD. Due to the geometry of the setup, we assume in the following $\Ud > U$. Importantly, since we only aim to detect the dynamics of the \textit{dot}, we simplify the description of the SQD by lifting its spin degree of freedom (e.g., with a magnetic field), such that it can be mostly occupied by a single electron with a fixed spin. However, an experimentally more realistic spin-degenerate SQD does not lead to significantly different measurement results if the SQD on-site interaction is large enough, see Sec.~\ref{sec_measurement_protocols} and Appendix~\ref{app_SDSQD}.

Electrons in the metallic leads are assumed to be noninteracting and are described by
\begin{eqnarray}
H_\text{leads} & = & \sum_{k,\sigma,\alpha=\mrm{L,R,C}}\epsilon_{k,\sigma,\alpha}c^{\dagger}_{k,\sigma,\alpha}c^{}_{k,\sigma,\alpha}\ ,
\end{eqnarray}
with creation (annihilation) operators $c^{\dagger}_{k,\sigma,\alpha}(c^{}_{k,\sigma,\alpha})$ for  an electron with spin $\sigma$ and momentum $k$ in lead $\alpha=\mrm{L,R,C}$. 
Each of the leads is in equilibrium, characterized by the chemical potential  $\mu_\alpha$ and the temperature $T$. Without loss of generality, we measure all energies with respect to $\mu_\mrm{C} = 0$.  
The dot is tunnel coupled to the central lead, $\alpha=\mrm{C}$,
with 
a tunneling matrix element $V_\mrm{C}$ that is assumed to be spin- and momentum independent; similarly, the detector dot couples to the left (L) and right (R) leads
with 
tunnel matrix elements $V_\mrm{L}$ and $V_\mrm{R}$,
\begin{eqnarray}
H_\text{tun} & = &\sum_{k,\alpha=\mrm{L,R}}\left[V_{\alpha}c^{\dagger}_{k,\sigma,\alpha}d^{}_\mrm{SQD}+\mrm{H.c.} \right]\nonumber\\
&&+\sum_{k,\sigma}\left[V_\mrm{C}c^{\dagger}_{k,\sigma,\mrm{C}}d^{}_{\mrm{d},\sigma}+\mrm{H.c.} \right].
\end{eqnarray}
\end{subequations}
The tunnel couplings are characterized by the spin-independent tunnel coupling strengths $\Gamma_\alpha=2\pi\rho_\alpha\left|V_\alpha\right|^2$, with the density of states $\rho_\alpha$ of the leads. We work in the wide-band limit and assume  $\Gamma_\alpha$ and $\rho_\alpha$ to be energy-independent. For convenience, we also define the total SQD tunneling strength $\Gamma = \GamL + \GamR$.
To be able to measure the dynamics of the dot, the characteristic time scale for tunneling processes between the SQD and the leads L,R must be much smaller than for tunneling events between the dot and the reservoir C. When considering explicit measuring protocols, we will therefore require $\GamC/\Gamma \ll 1$, which is the so-called measurement limit.

\subsection{Relaxation after a switch} \label{sec_relaxation}

\subsubsection{Generalized master equation}\label{sec_relaxation_mastereq}
To describe the relaxation behavior of the full d-SQD system brought out of equilibrium at a certain time $t=t_0$, we employ the techniques  of Refs.~\onlinecite{Splett10,Contreras12}, which we describe only briefly here. The time evolution of the d-SQD system after a switch obeys the {generalized master equation}
\begin{equation}\label{eq_kinetic}
 \ddt{\boldsymbol{P}(t)} = \int_{t_0}^tdt'\boldsymbol{W}(t,t')\cdot\boldsymbol{P}(t')\ .
\end{equation}
The vector $\boldsymbol{P}(t)$ contains the time-dependent occupation probabilities in the eigenstates of the isolated d-SQD system
\begin{equation}\label{eq_prob_vector}
 \boldsymbol{P} = \left(P_{00},P_{0\uparrow},P_{0\downarrow},P_{02},P_{10},P_{1\uparrow},P_{1\downarrow},P_{12}\right)^T\ .
\end{equation}
The first subscript indicates whether the SQD is  occupied (1) or unoccupied (0); the second subscript stands for the state of the quantum dot (d), which can be empty (0), singly occupied with a spin-up ($\uparrow$) or spin-down ($\downarrow$), or doubly occupied (2). The kernel $\boldsymbol{W}(t,t')$ is an $8\times 8$ matrix that accounts for transitions between the eigenstates of the d-SQD system via tunneling to the leads. \footnote{Note that for the d-SQD system, weakly tunnel-coupled to its environment as considered in this work, the kinetic equations for the occupation probabilities and the coherences decouple. Since we are interested in the occupation probabilities and the SQD current, we only consider the kinetic equation for the probabilities $\POft$.}
As the Hamiltonian is no longer time-dependent after the switch at time $t_0$, $\mat{W}$  depends only on the time difference $t-t'$ for times $t,t' > t_0$, i.e.,  $\boldsymbol{W}(t,t') = \boldsymbol{W}(t-t')$. 

To study the exponential relaxation, it is enough to consider times $t$ such that $t-t_0$ is much larger than the support of the kernel $\mat{W}(t-t')$, which for weakly coupled systems is typically of the order of $\hbar/\kBT$. In this long-time limit, we can approximate the kinetic equation, Eq.~\eqref{eq_kinetic}, as 
\begin{equation}\label{eq_kinetic_laplace}
  \ddt{\POft} = \sum_{n=0}^\infty\frac{1}{n!}\left[\frac{d^n\mat{W}(z)}{dz^n}\right]_{z=0}\frac{d^n\POft}{dt^n},
 \end{equation}
where $\mat{W}(z) = \int_{-\infty}^t dt'\mat{W}(t-t')e^{-z(t-t')}$ is the Laplace transform of the kernel.\par

Next, using the fact that we assume weak tunnel couplings, $\Gamma_\alpha/\kBT \ll 1$,\footnote{Strictly speaking, in the measurement limit, $\GamC/\Gamma \ll 1$, we cannot entirely exclude influences of higher order processes by assuming $\Gamma_\alpha/\kBT \ll 1$ alone, as higher powers of $\Gamma/\kBT$ can still be of the same order of magnitude as contributions linear in $\GamC/\kBT$. Consequently, we must also explicitly require $(\Gamma/\kBT)^2 \ll \GamC/\kBT$.} we expand  Eq.~\eqref{eq_kinetic_laplace} in powers of the tunnel coupling strengths $\Gamma_\alpha$. In first order in $\Gamma_\alpha$, the master equation predicts $\POft$ to decay purely exponentially,
\begin{equation}\label{eq_master_full}
 \POft = \exp{\left(\mat{A}t\right)}\cdot\Pin,
\end{equation}
where $\Pin = \mbf{P}(t_0)$ is the  initial probability distribution  at time $t_0 \equiv0$. Up to the leading order in the tunnel couplings, the transition matrix $\mat{A}$ is simply the first order in $\Gamma_\alpha$ contribution to the zero-frequency Laplace transform of the kernel, that is $\mat{A} = \mat{W}^{(1)}(z=0)$ with system parameters taking their values after the switch.

The elements of the transition matrix $\mat{A}$ can be explicitly calculated with the help of Fermi's golden rule or a perturbative real-time diagrammatic approach.~\cite{Konig96a,Konig96b} For the problem considered here,  $\mat{A}$ has seven real negative eigenvalues $-\lambda_i > 0$ corresponding to the negatives of the seven relaxation rates $\lambda_i$ which govern the decay of $\POft$, and a single zero eigenvalue. This ensures that the system relaxes to the stationary state $\Pst = \limes{t}{\infty}\POft$, determined by $\mat{A}\cdot \Pst = 0$ and $e^T\cdot\Pst = 1$ with the trace operator $e^T = (1,1,1,1,1,1,1,1)$.\par

In order to gain a physical understanding of the decay behavior, we consider the left and right eigenvectors of the non-Hermitian transition matrix $\mat{A}$. More explicitly, the decomposition of the vector $\boldsymbol{P}(t)$ in terms of the right eigenvectors $\boldsymbol{P}_i$ of the matrix $\boldsymbol{A}$ reveals the decay eigenmodes corresponding to the relaxation rates $\lambda_i$,
\begin{equation}
 \POft = \Pst + \sum_i(q_i^\mrm{in}-q_i^\mrm{st})\mbf{P}_ie^{-\lambda_i t}\label{eq_prob_vector_expansion}.
\end{equation}
The symbols $q_i^\mrm{in}$ and $q_i^\mrm{st}$ represent the initial and the stationary values of the time-dependent {quantities $q_i(t)$ that decay only with the single relaxation rate $\lambda_i$. These quantities can be directly related to the left eigenvectors $\hat{l}_i$ of the transition matrix $\mat{A}$ via $\hat{l}_i = \hat{q}_i - q_i^\mrm{st}e^\mrm{T}$, with an eight-dimensional row vector $\hat{q}_i$ defined to yield $q_i(t)$ via
\begin{equation}\label{eq_quantity}
 q_i(t) = \hat{q}_i\cdot\POft.
\end{equation}

\subsubsection{Quantum-dot subsystem}
When studying the effect of the back-action of a measurement on the quantum dot, we are interested in the occupation probability vector obtained by tracing out the degrees of freedom of the SQD,
\begin{equation}\label{eq_dot_probvector}
 \Pd = \begin{pmatrix}P_\mrm{d,0} \\ P_\mrm{d,\uparrow} \\ P_\mrm{d,\downarrow} \\ P_\mrm{d,2}\end{pmatrix} = \begin{pmatrix}P_{00} + P_{10} \\ P_{0\uparrow} + P_{1\uparrow} \\ P_{0\downarrow} + P_{1\downarrow} \\ P_{02} + P_{12}\end{pmatrix}.
\end{equation}

\subsubsection{Measuring {the} current through the SQD}
In order to detect the dynamics of the quantum dot after a fast switch, we study its effects on the current flowing through the sensor dot, from lead L to lead R, between which a bias voltage is applied. In general, the current in lead $\alpha$ is defined as $I_\alpha(t) = -e\frac{d\langle N_\alpha\rangle(t)}{dt}$, with the charge number operator $N_\alpha$ for electrons in lead $\alpha$ and the electron charge $e$. This current is determined explicitly by calculating the current kernel~\cite{Konig96a,Konig96b} which only includes tunneling rates for processes transferring particles between the SQD and lead $\alpha$. Expanding this kernel in first order in $\Gamma_\alpha$ leads to 
\begin{equation}
 I_\alpha(t) = e\,\hat{\mrm{n}}_\mrm{SQD}\cdot\mat{A}_\alpha\cdot\POft.\label{FormalismTunnelCurrentFinal}
\end{equation}
The row vector $\hat{{n}}_\mrm{SQD} = (0,0,0,0,1,1,1,1)$ corresponds to the SQD charge operator in our model, the matrix $\mat{A}_\alpha$ includes all terms of the full transition matrix $\mat{A}$ that are related to tunneling with lead $\alpha$.\par

\section{Results}
\label{results}

We start this results section with a brief summary of the decay behavior of the quantum dot when the coupling to the sensor dot is switched off.~\cite{Splett10,Contreras12} Next, we investigate  the dynamics of the dot coupled to the SQD. Our main focus will be to understand back-action effects. In the last part of this section, we use the knowledge of the back-action to establish measurement protocols aimed at detecting the original dot relaxation rates even in the presence of back-action effects.

\subsection{Decay dynamics of a single-level quantum dot}\label{sec_singledot}

We consider the quantum dot tunnel coupled to the reservoir C but \textit{completely decoupled} from the SQD,  i.e. $\UC = 0$. 
\begin{widetext}
The exponential decay  $\mbf{P}_\mrm{d}(t) = \exp\left(\mat{A}_\mrm{d}t\right)\cdot\mbf{P}_\mrm{d}^\mrm{in}$ of the dot probability vector defined in Eq.~\eqref{eq_dot_probvector} is then governed by the dot transition matrix $\mat{A}_\mrm{d}$, which is explicitly given by
\begin{equation}\label{eq_dotTransitionMatrix}
 \mat{A}_\mrm{d} = \frac{\GamC}{\hbar}
 \begin{pmatrix}
 -2 \fC{+} &  \fC{-} &  \fC{-} &  0\\
 \fC{+} & -\fC{-}-\fCUd{+} &  0 &  \fCUd{-}\\
 \fC{+} &  0 & -\fC{-}-\fCUd{+} &  \fCUd{-}\\
 0 &  \fCUd{+} &  \fCUd{+} &  -2\fCUd{-}
\end{pmatrix}\ .
\end{equation}
\end{widetext}
The Fermi functions $f^+_\alpha(\epsilon) = (\mrm{exp}\left[\beta(\epsilon - \mu_\alpha)\right]+1)^{-1}$ and $ f^-_\alpha(\epsilon) =1 - f^+_\alpha(\epsilon) $ are defined for the leads $\alpha$, where $\beta = 1/\kBT$ is the inverse temperature.
By diagonalizing $\mat{A}_\mrm{d}$, the decay dynamics of the dot probability vector $\mbf{P}_\mrm{d}(t)$ are found to be completely characterized by three independent decay eigenmodes, each of which is associated with a single relaxation rate that exclusively governs the exponential decay of one of the following three observables: charge $n_\mrm{d}(t) = P_\mrm{d,\uparrow}(t) + P_\mrm{d,\downarrow}(t) + 2P_\mrm{d,2}(t)$, spin $\sigma(t) = P_\mrm{d,\uparrow}(t) - P_\mrm{d,\downarrow}(t)$, as well as a third quantity that can be explicitly written as
\begin{equation}\label{eq_fermion_parity_quantity}
 m(t) = p\cdot P_\mrm{d,0}(t) + (1-p)\cdot P_\mrm{d,2}(t)\ ,
\end{equation}
with $p = \fCUd{+}/(\fC{-}+\fCUd{+})$. This third quantity $m(t)$ is linked to the fermion parity of the quantum dot, since $\mbf{P}_\mrm{fp} = (1,-1,-1,1)^T$ constitutes the right eigenvector of this decay mode.

We label the relaxation rates corresponding to the three quantities as $\lambda_\mrm{c}$ for the charge $n_\mrm{d}(t)$, $\lambda_\sigma$ for the spin $\sigma(t)$, and $\lamm$ for $m(t)$. The rates read
\begin{subequations}
\begin{align}
 \lambda_\mrm{c} & = \frac{\GamC}{\hbar}\left[\fC{+} + \fCUd{-}\right]\label{eq_relaxrates_single_c}\\
 \lambda_\sigma & = \frac{\GamC}{\hbar}\left[\fC{-} + \fCUd{+}\right]\label{eq_relaxrates_single_sig}\\
 \lamm & = 2\frac{\GamC}{\hbar}\label{eq_relaxrates_single_fp}\ .
\end{align}
\label{eq_relaxrates_single}
\end{subequations}
The behavior of the rates as a function of $\epsd$ is displayed in Fig.~\ref{fig_dotmodel}. When the transition energies between different dot states are far above the Fermi level on the scale of the temperature, $\epsd/\kBT \gg 1$, or far below, $-(\epsd+\Ud)/\kBT \gg 1$, the charge and spin rate are simply determined by the bare tunnel coupling $\GamC/\hbar$. However, in the regime of single occupation, $-\Ud < \epsd < 0$, we find that the charge rate is enhanced to $\lamc \lesssim 2\GamC/\hbar$, whereas the spin rate is suppressed to almost zero, $\lamsig\gtrsim 0$. The reason for the deviation from $\GamC/\hbar$  in the region in which the dot is preferably singly occupied is the following: the spin-degeneracy requires a vanishing spin $\sigma^\mrm{st} = \sigma(t\rightarrow\infty) = 0$ in the stationary limit, but a singly occupied dot can never have spin zero, so that the \textit{spin 
average} $\sigma = P_\mrm{d,\uparrow} - P_\mrm{d,\downarrow}$ can only reach its stationary value $\sigma^\mrm{st} = 0$ by thermal fluctuations. Thus, the relaxation rate $\lamsig$ is strongly suppressed. On the other hand, the spin degree of freedom offers two possibilities, namely $\uparrow$ and $\downarrow$, to reach the stationary charge number $n_\mrm{d}^\mrm{st}$, thereby doubling  the charge relaxation rate. 

 \begin{figure}[t]
\includegraphics[width=\linewidth]{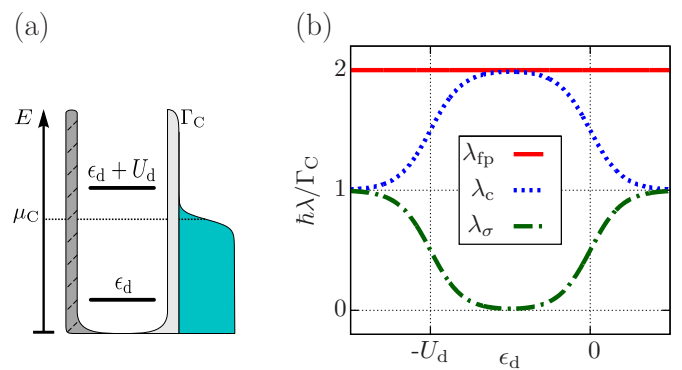}
\caption{
 \label{fig_dotmodel}
(a) Schematic representation of the dot (d) coupled to a reservoir (C). The dot is driven out of equilibrium at time $t_0$ and then relaxes to a new equilibrium state. (b) Relaxation rates that govern the exponential decay of the charge $(\lamc)$, the spin $(\lamsig)$ and the fermion-parity mode $(\lamm)$ of the dot as a function of the level position $\epsd$ after the switch. The temperature is set to $\kBT = \Ud/10$.}
 \end{figure}\par

The decay rate $\lamm = 2\GamC/\hbar$ of the fermion-parity mode is completely independent of temperature, interaction strength and level position, see Eq.~\eqref{eq_relaxrates_single_fp} and the red {solid} line in Fig.~\ref{fig_dotmodel}. Recently, it was demonstrated that regardless of how many orders in the tunnel couplings are included in the perturbative expansion, the fermion-parity operator of multi-orbital systems always constitutes a right eigenvector of the transition kernel as long as the wide band limit is considered.~\cite{Saptsov12,Saptsov13} The corresponding eigenvalue is the negative sum of all bare tunnel couplings that connect the reduced system to its environment. The reason for the simple form of the fermion-parity rate is the fact that the fermion-parity operator is not affected by any finite-temperature corrections arising as a consequence of the reservoir coupling.~\cite{Saptsov12,Saptsov13}  

To understand the role of $\lamm$ for the quantum dot state relaxation, let us examine in more detail the quantity $m(t)$, see Eq.~\eqref{eq_fermion_parity_quantity}. As a function of the level position $\epsd$, the coefficient $p $ entering $m(t)$ drops from $1$ to $0$ for increasing $\epsd$ around the particle-hole symmetric point $\epsd = -\Ud/2$, at which we find $p = 1/2$. The quantity $m(t)$ thus approximately equals the zero-occupation probability $m(t) \approx P_\mrm{d,0}(t)$ for final level positions $\epsd < -\Ud/2$ where the stationary zero-occupation probability $P_\mrm{d,0}^\mrm{st} \approx 0$ is close to zero, whereas it approximately equals the double-occupation probability $m(t) \approx P_\mrm{d,2}(t)$ if $\epsd > -\Ud/2$ and thus $P_\mrm{d,2}^\mrm{st} \approx 0$.\par

One way to excite the dynamics of $m(t)$, and to consequently let $\lamm$ contribute significantly to the dot state relaxation, is therefore to prepare the dot in the empty state in a regime where the dot in equilibrium would be doubly occupied; in this situation $m^\mathrm{excited}\approx P_{\mathrm{d},0}$ is close to $1$ directly after the switch. This can be realized by applying a gate pulse from an initial state at $\epsd>-\Ud/2$, where $P_\mrm{d,0}^\mrm{in} \neq 0$, to the regime $\epsd < -\Ud/2$. Vice versa, one could prepare the dot in the doubly occupied state in a region where the dot in equilibrium would be empty. In Sec.~\ref{sec_measurement_protocols}, we will indeed suggest a gate-pulse protocol based on the described behavior to detect the rate $\lamm$ independently from the charge relaxation rate $\lamc$.\par

\subsection{Master equation of the dot-sensor system}
\label{sec_dot-sensor}

To understand the dynamics of the full d-SQD system, we first discuss the \textit{structure} of the master equation  
$d\POft/dt =\mat{A}\cdot \POft$ with~\footnote{The sum over all matrix elements in one column of $\mat{A}$ vanishes due to probability conservation. Here, this rule even applies to the \textit{individual contributions} of the dot and the SQD because the dot/SQD degrees of freedom commute with the tunnel coupling of the SQD/dot to its leads, see related discussions in Refs. \onlinecite{Salmilehto12} and \onlinecite{Hell13}}
\begin{equation}
\mat{A} = \left(\begin{array}{cc}
\mat{A}_\mrm{d} &0\\
0 & \overline{\mat{A}}_\mrm{d}\end{array}\right)
+\left(\begin{array}{cc}
-\mat{A}^+_\mrm{SQD} & \mat{A}^-_\mrm{SQD}\\
\mat{A}^+_\mrm{SQD} & - \mat{A}^-_\mrm{SQD}
\end{array}\right).\label{eq_master_spinless}
\end{equation}
The upper diagonal block of the first matrix describes tunneling processes between the quantum dot  and the tunnel-coupled reservoir C when the SQD is completely empty. It is given by the $4\times 4$ matrix $\mat{A}_\mrm{d}$,  Eq.~\eqref{eq_dotTransitionMatrix},  governing the dynamics of the quantum dot in the absence of a detector. In the lower right diagonal block of the first matrix, we find the transition matrix $\overline{\mat{A}}_\mrm{d}$, which differs from $\mat{A}_\mrm{d}$ by an overall energy shift $\epsd \rightarrow \epsd + \UC$ accounting for the capacitive coupling to an occupied SQD. Any energy shift by $+\UC$ is indicated by barred symbols, such that $\bar{a}(\epsilon) := a(\epsilon+\UC),\quad \bar{\bar{a}}(\epsilon) := a(\epsilon+2\UC)$ for an arbitrary energy-dependent object $a(\epsilon)$.\par

The  $4\times 4$ blocks {$\mat{A}^\pm_\mrm{SQD}$} of the second matrix represent electron hopping from the leads L and R to the SQD $(+)$ and vice versa $(-)$. They are given by the diagonal matrices
\begin{equation}
 \mat{A}^\pm_\mrm{SQD} = \frac{\Gamma}{\hbar}\cdot\mrm{diag}\left(\fS{\pm}, \bfS{\pm}, \bfS{\pm}, \bbfS{\pm}\right)\ .
\end{equation}
The first element of $\mat{A}^\pm_\mrm{SQD}$ is the SQD tunneling rate for an empty dot. Its energy dependence is determined by
\begin{equation}
 \fS{\pm} = \sum_{\alpha=\mrm{L,R}}\frac{\Gamma_\alpha}{\Gamma}\falp{\pm},\quad \Gamma = \GamL + \GamR\ .
\end{equation}
The second and third element correspond to the same transition, but with the dot singly occupied; the fourth element represents the case of a doubly occupied dot. All off-diagonal elements of $\mat{A}^\pm_\mrm{SQD}$ vanish because they refer to simultaneous tunneling processes in SQD and dot which do not contribute in the leading order in the tunnel couplings.

\subsection{Decay dynamics during measurement}\label{sec_dynamics_measurement}

We first transform the master equation from the energy eigenbasis, Eq.~\eqref{eq_master_spinless}, into a new basis that directly reveals the time evolution of the dot quantities of interest, namely the spin $\sigma(t)$, the dot charge $n_\mrm{d}(t)$ and the fermion-parity quantity $m(t)$. The full master equation thereby assumes the block-diagonal form 
$$\mat{A}\rightarrow \left(\begin{array}{cccc}0 &&&\\ &-\lambda_\mathrm{gfp} & &\\ && -\mat{A}_\sigma&\\ &&&-\mat{A}_\mathrm{c}\end{array}\right), $$ 
where the zero eigenvalue is simply related to the existence of a stationary state. The other three blocks describe the decay of the \textit{global} fermion-parity mode for the d-SQD system ($\lambda_\mathrm{gfp}$), the dot spin ($\mat{A}_\sigma$), and the charge as well as the \textit{local} dot fermion-parity mode ($\mat{A}_\mathrm{c}$).\footnote{The local SQD fermion-parity mode is not mentioned here, since in the spinless case, it \textit{is given} by the local SQD charge decay mode.} In the following, we discuss each of the three blocks in more details and provide the definition of $\lambda_\mathrm{gfp}$, $\mat{A}_\sigma$, and $\mat{A}_\mathrm{c}$.

\subsubsection{Global fermion-parity mode}\label{sec_global_fp}
The right eigenvector of the transition matrix $\mat{A}$, which describes the global fermion-parity eigenmode of the d-SQD system contains in each component the fermion parity of the corresponding many-particle eigenstate in the reduced system:
\begin{equation}
 \mbf{P}_\mrm{gfp} = (1,-1,-1,1,-1,1,1,-1)^T\ .\label{eq_fermion_parity_vector} 
\end{equation}
The associated relaxation rate is given by the sum of all bare tunnel couplings, yielding
\begin{equation}
 \lambda_{\text{gfp}} = \frac{1}{\hbar}\left(\Gamma + 2\GamC\right)\ .
\end{equation}
From the eigenmode expansion \eqref{eq_prob_vector_expansion} and the form of the eigenvector Eq.~\eqref{eq_fermion_parity_vector}, it is easy to prove that the  global  fermion-parity mode cannot enter the time evolution of any quantity  local  either to the dot or the SQD.
It is, in particular, irrelevant for the measurement of the quantum dot decay behavior via the sensor dot current, as this current   
only depends on local quantities, see Sec.~\ref{sec_measurement}.

\subsubsection{Spin sector}\label{sec_spinsector}
Since the SQD is only capacitively coupled to the dot, it does not measure the dot spin. However, by shifting the dot addition energy, the SQD still has an effect on the spin relaxation, giving rise to back-action without measurement. Introducing the two \textit{conditional spin variables}
\begin{equation}
 \sigma_0(t) = P_{0\uparrow}(t) - P_{0\downarrow}(t),\quad \sigma_1(t) = P_{1\uparrow}(t) - P_{1\downarrow}(t),\ 
\end{equation}
which constitute the dot spin under the condition that the SQD is either empty $(0)$ or filled $(1)$, one finds
\begin{equation}
 \ddt{}\begin{pmatrix}\vphantom{\frac{\displaystyle E}{\displaystyle E}}\sigma_0(t) \\ \vphantom{\frac{\displaystyle E}{\displaystyle E}}\sigma_1(t)\end{pmatrix} = -\underbrace{\begin{pmatrix}\vphantom{\frac{\displaystyle E}{\displaystyle E}}\lamsig + \frac{\Gamma}{\hbar}\bfS{+} & \vphantom{\frac{\displaystyle E}{\displaystyle E}}-\frac{\Gamma}{\hbar}\bfS{-}\\ \vphantom{\frac{\displaystyle E}{\displaystyle E}} -\frac{\Gamma}{\hbar}\bfS{+} & \vphantom{\frac{\displaystyle E}{\displaystyle E}} \blamsig + \frac{\Gamma}{\hbar}\bfS{-} \end{pmatrix}}_{\mat{A}_\sigma :=}\cdot\begin{pmatrix}\vphantom{\frac{\displaystyle E}{\displaystyle E}}\sigma_0(t) \\ \vphantom{\frac{\displaystyle E}{\displaystyle E}}\sigma_1(t)\end{pmatrix}\label{eq_spin_master}\ .
\end{equation}
The transition matrix $\mat{A}_\sigma$ contains the spin relaxation rate $\lamsig$ of the dot in the absence of the SQD given in Eq.~\eqref{eq_relaxrates_single_sig}, its energy-shifted version $\blamsig$, as well as the contributions $\bfS{\pm}$ from the SQD occupation.\par

The dynamics of the dot spin, $\sigma(t) = \sigma_0(t) + \sigma_1(t)$, is obtained by tracing out the states of the SQD, yielding
\begin{equation}\label{eq_spineq}
 \ddt{\sigma(t)} = -\lamsig\sigma_0(t) - \blamsig\sigma_1(t)\ .
\end{equation}
If the SQD is empty during the entire time evolution, the spin decay is only described by the first term on the r.h.s. of Eq.~(\ref{eq_spineq}) corresponding to the already known spin rate $\lamsig$. The second term instead accounts for the case when the SQD is constantly filled, causing a capacitive energy shift in the dot and thus modifying the spin decay rate as $\lamsig \rightarrow \blamsig$. Hence, back-action here leads to a capacitive shift of decay rates of the dot. 

\begin{figure}
 \includegraphics[width=\linewidth]{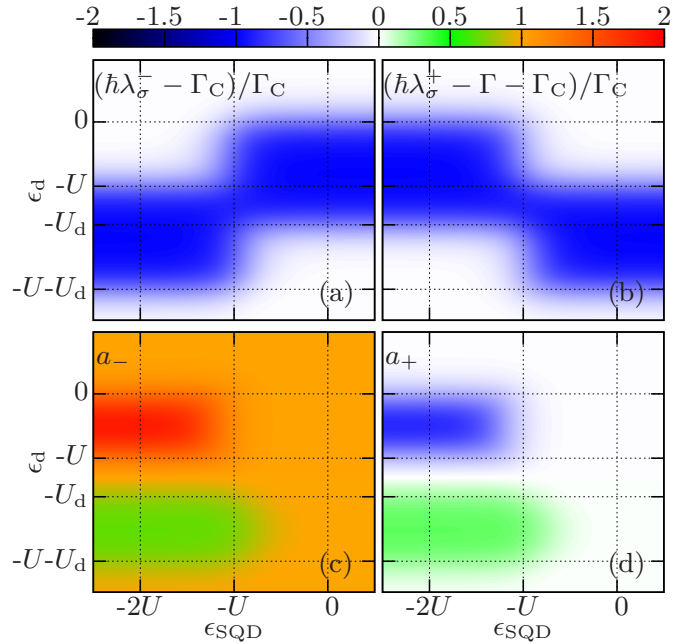}
 \caption{(a),(b): Deviation of the spin relaxation rates $\lamsigpm$ from their values in the noninteracting limit $U,\Ud\rightarrow 0$. (c),(d): The coefficients $a_\pm$ entering the time-dependent spin decay expressed in Eq.~\eqref{eq_spin_decay}, for a switch applied to a system in which the quantum dot is initially occupied with a spin-up electron and the SQD is initially empty. All quantities are plotted as a function of the final level positions $\epsSQD$ and $\epsd$. The other system parameters are: $\mu_\mrm{L} = \mu_\mrm{R} = \mu_\mrm{C} = 0$, $\GamL = \GamR = \GamC = \Gamma/2$, $\Ud = 8U/5$, $\kBT = \Ud/10 = 0.16U$, $\Gamma/k_\mathrm{B}T\ll1$.}
 \label{fig_spin}
\end{figure}\par

However, the probabilities for the SQD to be filled and to be empty can simultaneously be nonzero and can also evolve in time. The total spin time evolution is therefore influenced by a combination of both contributions in the differential equation~\eqref{eq_spineq}. Consequently, the dot spin is governed by two relaxation rates, obtained by diagonalizing the transition matrix $\mat{A}_\sigma$. In other words, back-action does not only modify rates, it also leads to an onset of new decay modes. With $\sigma^\mrm{st} = \sigma(t\rightarrow\infty) = 0$, the most general form which $\sigma(t)$ can assume is
 \begin{equation}
  \sigma(t) = a_-e^{-\lamsigm t} + a_+e^{-\lamsigp t}\label{eq_spin_decay}\ .
 \end{equation}
The relaxation rates corresponding to these decay modes are given by
 \begin{eqnarray}
  \lamsigpm & = & \frac{1}{2}\left[\frac{\Gamma}{\hbar} + \lamsig + \blamsig \right. \label{eq_spin_rates}\\
  & & \pm \left.\sqrt{\left(\frac{\Gamma}{\hbar} - \lamsig + \blamsig\right)^2 + \frac{4\Gamma}{\hbar}\bfS{+}(\lamsig - \blamsig)}\right]\notag ,
 \end{eqnarray}
representing a combination of the original dot spin rate $\lamsig$ and its capacitively shifted counter-part $\blamsig$.
The dependence of $\lamsigpm$ on both $\epsd$ and $\epsSQD$ is shown in Figs.~\ref{fig_spin}(a),(b). To emphasize back-action effects, we display for both rates their deviation from the noninteracting limits, $U=\Ud=0$, where $\lamsigm \rightarrow \GamC/\hbar$ and $\lamsigp \rightarrow (\Gamma + \GamC)/\hbar$. We consider an equilibrium situation $\mu_\mrm{L} = \mu_\mrm{R} = \mu_\mrm{C} = 0$ and symmetric coupling, $\GamL = \GamR = \GamC = \Gamma/2$.~\footnote{As we are not interested in spin measurements, we here deviate from the measurement limit $\GamC/\Gamma \ll 1$. }

The rate  $\lambda_\sigma^-$ exactly reflects the context described for Eq.~\eqref{eq_spin_decay}. It becomes the original spin rate $\lamsig$ for $\epsSQD > -U$, and the capacitively shifted rate $\blamsig$ for level positions $\epsSQD < -U$ allowing the SQD to be stably occupied in the presence of a single dot electron. The relaxation rate $\lambda_\sigma^+$ behaves, apart from an overall shift by $\Gamma/\hbar$, almost mirrored compared to $\lamsigm$ as a function of $\epsSQD$. In the transitional regime $\epsSQD \approx-U$, temperature broadening leads to a combination of the spin relaxation rates $\blamsig$ and $\lamsig$ for both $\lamsigp$ and $\lamsigm$.\par

To gain a better physical understanding of the two rates, we study the coefficients $a_\pm$ yielding the relative contribution of each exponential function governing the spin decay, Eq.~\eqref{eq_spin_decay}, and thus measuring the relevance of the relaxation rates $\lamsigpm$ for the spin decay $\sigma(t)$ for given initial conditions as a function of the final level positions $\epsd,\epsSQD$. Using the left and right eigenvectors $\hat{l}^\pm_{\sigma}$ and $\mbf{P}^\pm_{\sigma}$ of the full transition matrix $\mat{A}$ corresponding to the relaxation rates $\lamsigpm$, see Appendix \ref{app_spinvec}, the coefficients are obtained by
 \begin{equation}
  a_\pm = \left(\hat{\sigma}\cdot\mbf{P}^\pm_{\sigma}\right)\left(\hat{l}^\pm_{\sigma}\cdot\Pin\right)\ .
 \end{equation}
The row vector $\hat{\sigma} = (0,1,-1,0,0,1,-1,0)$ is the vector representation of the spin operator.\par

In Figs.~\ref{fig_spin}(c),(d), we display $a_\pm$ as a function of the values $\epsd$ and $\epsSQD$ after the switch, assuming that initially, the dot is occupied by a spin-up electron and the SQD is empty. The behavior of  $a_-$ reveals that the relaxation dynamics are mostly dominated by the relaxation rate $\lambda_\sigma^-$. In particular, it is the only contributing rate as long as the switch does not induce any time evolution of the SQD state. 
However, if the latter is the case and at the same time the dot state is susceptible to the SQD state, $a_+$ indicates that $\lambda_\sigma^+$ also contributes to the decay of the dot spin exactly for those level positions, $-U < \epsd < 0$ and $-U < \epsd + \Ud < 0$. 
Yet, we find this influence to decrease with the ratio $\GamC/\Gamma$, and even to be negligible in the measurement limit  $\GamC/\Gamma \ll 1$. Note that in the measurement limit, all contributions proportional to  $e^{-\lamsigp t}$ also decay on a much smaller time scale compared to those proportional to $e^{-\lamsigm t}$.\par

We have performed further analyses of $a_\pm$ for initial conditions different from those chosen above, but they all lead to the same conclusion that, as long as the detector state is not affected by the gate switch, the spin still relaxes only at a single effective spin rate $\lamsigm$, taking into account a possible energy shift due to the capacitive coupling between the dot and the SQD.

\subsubsection{Charge sector}\label{sec_charge}
To study the decay dynamics of the dot charge $n_\mrm{d}(t)$ and the dot fermion-parity quantity $m(t)$, it is useful to analyze the following four quantities:
\begin{subequations}
\label{eq_charge_quantities}
\begin{eqnarray}
 n_\mrm{d,0}(t) & = & P_{0\uparrow}(t) + P_{0\downarrow}(t) + 2P_{02}(t),\\
 n_\mrm{d,1}(t) & = & P_{1\uparrow}(t) + P_{1\downarrow}(t) + 2P_{12}(t),\\
 n_\mrm{SQD}(t) & = & P_{10}(t) + P_{1\uparrow}(t) + P_{1\downarrow}(t) + P_{12}(t),\\
 P_\mrm{d,2}(t) & = & P_{02}(t) + P_{12}(t)\ .
\end{eqnarray}
\end{subequations}
The conditional dot charges, $n_\mrm{d,0/1}$, yield the time-dependent average dot charge number for either an empty $(0)$ or filled $(1)$ SQD, the time-dependent SQD charge number is labeled $n_\mrm{SQD}(t)$, and we have furthermore selected the double occupation probability $P_\mrm{d,2}(t)$, which together with $n_\mrm{d,0/1}$ can be used to calculate $m(t)$.
These four quantities generally have non-vanishing expectation values in the stationary state. Therefore, we study the set of differential equations for the deviation $\Delta q(t) = q(t) - q^\mrm{st}$ of each quantity $q$ in Eq.~\eqref{eq_charge_quantities} from its respective stationary limit $q^\mrm{st}$, namely, 
\begin{widetext}
\begin{equation}
 \ddt{}\begin{pmatrix}\Delta n_\mrm{d,0}(t) \\ \Delta n_\mrm{d,1}(t) \\ \Delta n_\mrm{SQD}(t) \\ \Delta P_\mrm{d,2}(t) \end{pmatrix} = -\underbrace{\begin{pmatrix}  \lamc + \frac{\Gamma}{\hbar}\bfS{+} &  -\frac{\Gamma}{\hbar}\bfS{-} &  2\frac{\GamC}{\hbar}\fd{+} &  -2\frac{\Gamma}{\hbar}\bDfS \\  -\frac{\Gamma}{\hbar}\bfS{+} &  \blamc + \frac{\Gamma}{\hbar}\bfS{-} &  -2\frac{\GamC}{\hbar}\bfd{+} &  2\frac{\Gamma}{\hbar}\bDfS \\  \frac{\Gamma}{\hbar}\DfS &  \frac{\Gamma}{\hbar}\DfS &  \frac{\Gamma}{\hbar} &  -\frac{\Gamma}{\hbar}(\DfS - \bDfS) \\  -\frac{\GamC}{\hbar}\fdUd{+} &  -\frac{\GamC}{\hbar}\bfdUd{+} &  0 &  2\frac{\GamC}{\hbar} \end{pmatrix}}_{\mat{A}_\mrm{c}:=}\cdot\begin{pmatrix}\Delta n_\mrm{d,0}(t) \\ \Delta n_\mrm{d,1}(t) \\ \Delta n_\mrm{SQD}(t) \\ \Delta P_\mrm{d,2}(t) \end{pmatrix}\ ,\label{eq_chargeblock}
\end{equation}
\end{widetext}
introducing the abbreviations
\begin{eqnarray*}
 &\fd{\pm} = \fC{\pm}, \quad \fdUd{\pm} = \fCUd{\pm},\\
 &\Dfd = \fd{+} - \bfd{+}, \quad \DfdUd = \fdUd{+} - \bfdUd{+},\\
 &\delta\! f_{\alpha = \mrm{L,R}} = \falp{+} - \bfalp{+},\\
 &\delta\! \bar{f}_{\alpha = \mrm{L,R}} = \bfalp{+} - \bbfalp{+},\\
  &\DfS = \sum_{\alpha=\mrm{L,R}}\frac{\Gamma_\alpha}{\Gamma}\delta\!f_\alpha, \quad \bDfS = \sum_{\alpha=\mrm{L,R}}\frac{\Gamma_\alpha}{\Gamma}\delta\!\bar{f}_\alpha\ .
\end{eqnarray*}
To get an overview over the physics involved in Eq.~\eqref{eq_chargeblock}, we first extract the differential equations for the dot charge $n_\mrm{d}(t)$ and $m(t)$. Subsequently, we discuss the different relaxation rates which enter the time-dependent decay of both quantities.\par

\begin{figure*}
\begin{minipage}[t]{0.32\linewidth}
\vspace{0pt}
\flushleft
\large (a)
\normalsize
\vspace{0pt}
 \includegraphics[width=\linewidth]{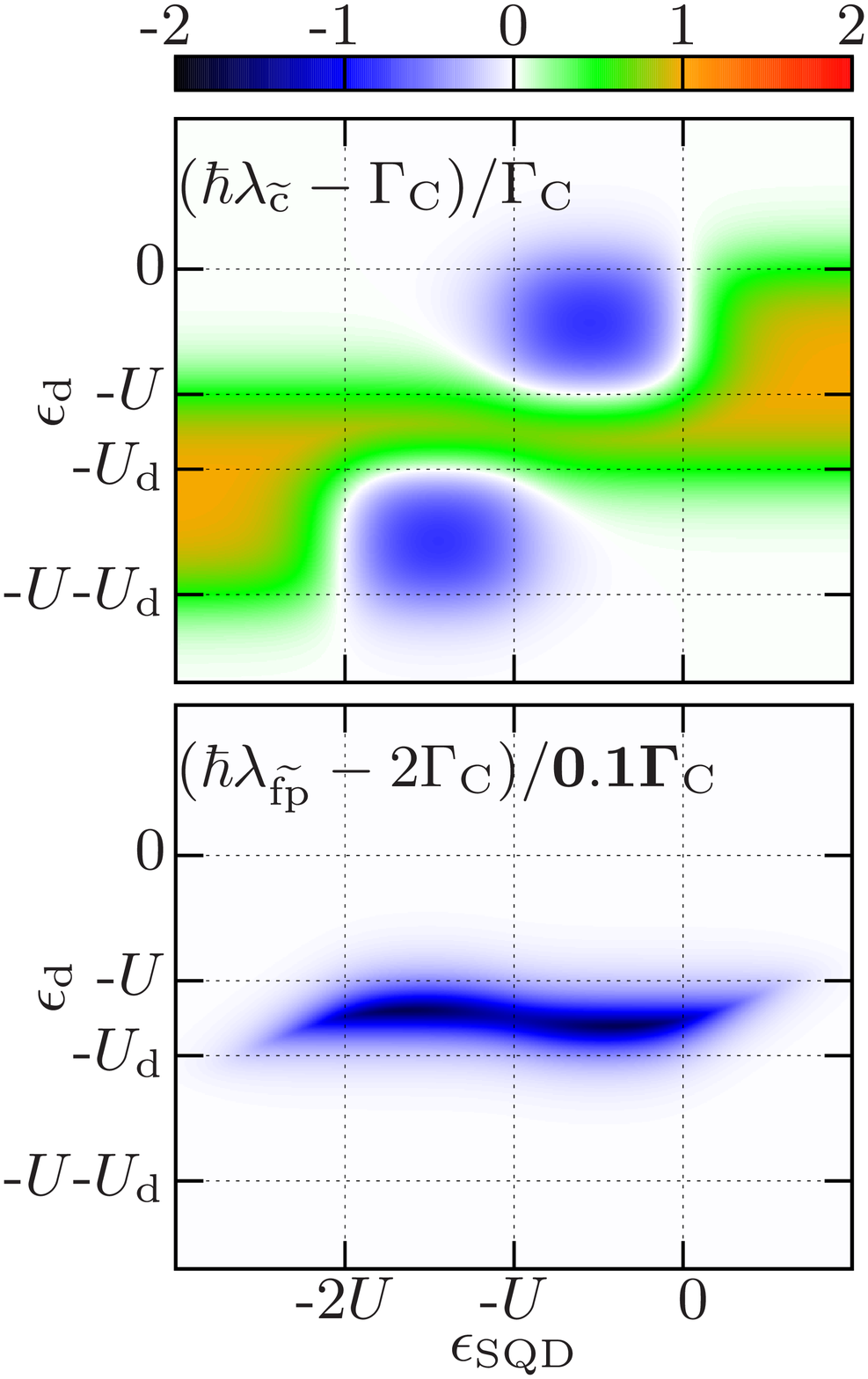}
\end{minipage}
\hfill
\begin{minipage}[t]{0.62\linewidth}
\vspace{0pt}
\flushleft
\large (b)
\normalsize
  \includegraphics[width=\linewidth]{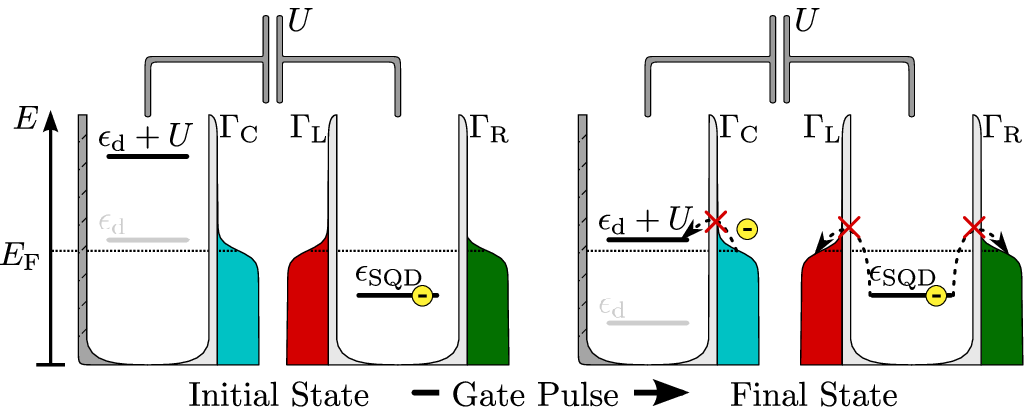}\\
\vspace{0.1cm}
\large (c)\\
\vspace{0.2cm}
\normalsize
  \includegraphics[width=\linewidth]{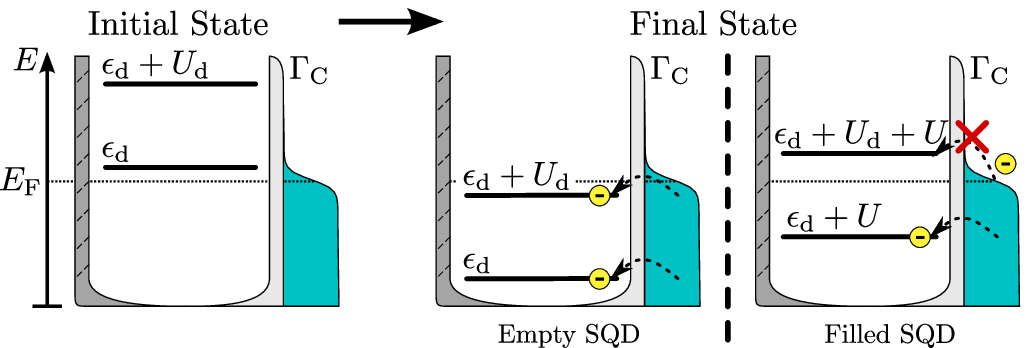}
\end{minipage}
  \caption{(a) Deviation of the dot related relaxation rates $\lamcm,\lamtm$ from their respective values in the noninteracting limit $U,\Ud\rightarrow 0$ as a function of the final level positions $\epsSQD$ and $\epsd$. We set $\mu_\mrm{L} = \mu_\mrm{R} = \mu_\mrm{C} = E_\mrm{F} = 0$, $\GamL = \GamR = \Gamma/2$, $\Ud = 8U/5$, $\kBT = \Ud/10 = 0.16U$. With $\GamC   = \Gamma/10$, we must also assume $\Gamma/\kBT \ll 0.1$, even though this ratio does not enter explicitly in our leading order calculations. (b) Typical gate pulse into the back-action regime $-U < \epsd < \epsSQD < 0$ for which the dot charge relaxation is strongly suppressed. The initially filled SQD induces Coulomb blockade in the dot, and the latter therefore cannot reach its stationary limit, which would be the singly occupied state. Note that for simplicity, the graph does not indicate the dot on-site repulsion strength $\Ud$. (c) An initially empty dot gate-pulsed into the regime $-\Ud - U < \epsd < -\Ud$. The dot is driven into double 
occupation only if the SQD is empty, but at least one electron can always enter the dot as long as $\Ud > U$, even 
if the 
SQD is 
occupied. }
  \label{fig_pseudospin}
\end{figure*}

The dot charge, $n_\mrm{d} = n_{\mrm{d},0} + n_{\mrm{d},1}$, obeys
\begin{eqnarray}\label{eq_chargeeq}
 \ddt{\Delta n_\mrm{d}(t)} & = & -\lamc\Delta n_\mrm{d,0}(t) -\blamc\Delta n_\mrm{d,1}(t) \notag\\
 & & -\frac{2\GamC\Dfd}{\hbar}\Delta n_\mrm{SQD}(t) .
\end{eqnarray}
The first two terms can be interpreted analogously to the time evolution of the dot spin, Eq.~\eqref{eq_spineq}. If the SQD is constantly empty, the dot charge decays mainly at the charge rate $\lamc$; if the SQD is singly occupied, the relaxation is primarily governed by the energy-shifted rate $\blamc$. However, in addition to these implicit dependencies on the SQD state, the last term in Eq.~\eqref{eq_chargeeq} also couples the charge dynamics {directly} to the SQD dynamics. Importantly, the structure of Eq.~(\ref{eq_chargeblock}) furthermore indicates that the time evolution of the dot charge depends on $m(t)$ via coupling of $n_\mrm{d,0}(t), n_\mrm{d,1}(t)$ and $ n_\mrm{SQD}(t)$ to the double-occupation probability. 

Indeed, the time evolution of $m(t)$ in the presence of a capacitively coupled SQD is described by 
\begin{eqnarray}\label{eq_master_fermionparity}
 \ddt{\Delta m(t)} & = &  - \frac{2\GamC}{\hbar}\Delta m(t) + \frac{2\GamC\Dfd\fdUd{+}}{\hbar\left[\fd{-} + \fdUd{+}\right]}\Delta n_\mrm{SQD}(t) \notag\\
  && - \frac{\GamC\left[\bfd{-}\fdUd{+}-\fd{-}\bfdUd{+}\right]}{\hbar\left[\fd{-} + \fdUd{+}\right]}\Delta n_\mrm{d,1}(t).\notag\\
\end{eqnarray}
Interestingly, if one neglects finite temperature broadening and assumes $\Ud > U$, the r.h.s. of Eq.~(\ref{eq_master_fermionparity}) becomes  $- \frac{2\GamC}{\hbar}\Delta m(t)$, implying that   $m(t)$ decays exponentially with a rate $\lamm = 2\GamC/\hbar$, independent of the dot and SQD charge state.\footnote{In the infinite temperature limit $T\rightarrow\infty$, this is true regardless of whether $\Ud > U$ or $\Ud < U$, but this regime is irrelevant for measurements.} 

In general, there are four relaxation rates, $\lamcm$, $\lamtm$, $\lamSET$ and $\lamcp$, obtained from diagonalizing the $4\times 4$ block $\mat{A}_\mrm{c}$, which influence the decay of {$\Delta n_\mrm{d}(t)$, $\Delta m(t)$ and $\Delta n_\mrm{SQD}(t)$ via}
 \begin{equation} \label{eq_charge_decay}
  \Delta q(t) = \sum_{i} a_i(q)e^{-\lambda_i t} ,
 \end{equation}
with $i\in \{\tilde{\text{c}}\ , \widetilde{\text{fp}}\ , \widetilde{\text{SQD}}\ , +\}$. Equivalently to the prefactors $a_\pm$ appearing in the spin decay, Eq.~\eqref{eq_spin_decay}, the coefficients $a_i(q)$ describe the relative influence of the four relaxation rates $\lambda_i$ on an observable $q$. A detailed analysis of these coefficients $a_i$ leads to the following conclusions:\cite{SchulenborgMasterThesis}
\begin{itemize}
 \item One of the rates, hence termed the \textit{effective charge relaxation rate} $\lamcm$, in many parameter regimes dominates the decay dynamics of the dot charge 
 $n_\mrm{d}(t)$.
 
 \item We furthermore identify an \textit{effective dot fermion-parity rate}, which we have already predicted from Eq.~\eqref{eq_master_fermionparity} to be the major influence on $m(t)$,  $\lamtm$.
 
 \item In many parameter regimes, one of the rates acts as an \textit{effective SQD charge relaxation rate}, $\lamSET$, that primarily governs the exponential decay of $n_\mrm{SQD}(t)$.
 
 \item Finally, there is a rate $\lamcp$, here called \textit{combined dot-SQD rate}, which has the property to be of importance for the decay of $n_\mrm{d}(t)$ and $m(t)$ if the SQD has a time-dependent influence on the dot, very similar to how $\lamsigp$ enters the spin decay, see Sec.~\ref{sec_spinsector}.
\end{itemize}
There are nevertheless also regimes of strong back-action, in which the separation of the role of these rates is no longer as clear. To elaborate on this, let us now discuss in more detail the gate dependence  of the rates. In Fig.~\ref{fig_pseudospin}(a), we show $\lamcm$ and $\lamtm$ as a function of the final level positions $\epsSQD$ and $\epsd$. Since we are interested in the measurement limit ($\GamC \ll \Gamma$),  the decay rates $\lamSET$ and $\lamcp$, both of the order of $\Gamma$, govern decay eigenmodes that die out much faster than the decay modes corresponding to $\lamcm$ and $\lamtm$. So even though $\lamSET$ and $\lamcp$ can in principle enter $n_\mrm{d}(t)$ and $m(t)$, their effect on the dynamics is mostly negligible on the dot time scale $\hbar/\GamC$, and thus will not be discussed in detail here.\par

The $\epsd$-dependence of the effective charge rate $\lamcm$ equals the one of the unmodified charge rate $\lamc$ for SQD levels $\epsSQD > 0$, and of the energy-shifted rate $\blamc$ for $\epsSQD < -2U$. This reflects the capacitive potential shift we have already encountered in the spin dynamics, and which is predicted by the first two terms in the differential equation of the dot charge, Eq.~\eqref{eq_chargeeq}. The effective charge rate $\lamcm$ additionally exhibits regions of suppressed relaxation for $-U < \epsd,\epsSQD < 0$ and $-U < \epsd+\Ud,\epsSQD + U < 0$. In these regions, also the impact of the other rates entering the charge $n_\mathrm{d}(t)$, especially the effective fermion-parity rate $\lamtm$, is enhanced.
The physical origin of the suppression of $\lamcm$ in these regimes of strong back-action is illustrated in Fig.~\ref{fig_pseudospin}(b), considering as an example a gate pulse that drives the system from an initially empty dot and filled SQD into the region $-U < \epsd < \epsSQD < 0$. The stationary state $\Pst$ for these level positions predicts the SQD to be empty and the dot to be singly occupied. However,  the required tunneling processes from the reservoir C to the dot as well as from the SQD to the source and drain leads are blocked by the Coulomb interaction between the two subsystems, and therefore the charge relaxation rate is strongly suppressed.\par

An equivalent argument for the suppression of $\lamcm$ holds for the second regime of strong back-action $-U < \epsd+\Ud < \epsSQD + U < 0$. However, transitions from an empty dot into this regime do not immediately drive the dot into Coulomb blockade, which is established only after one electron has entered the dot. The time evolution of $n_\mrm{d}(t)$ is therefore not completely suppressed, meaning that the other relaxation rates, in particular the fermion-parity rate $\lamtm$, also contribute to the exponential charge decay in this case. An analogous effect is observed when gate-pulsing the system from a doubly occupied state to $-U < \epsSQD < \epsd < 0$.\par

In principle, Coulomb blockade induced by the SQD also leads to a suppression of the dot fermion-parity rate $\lamtm$, as displayed in Fig.~\ref{fig_pseudospin}(a). However, we can also understand why this suppression is much weaker compared to what we find for the charge rate, and why $\lamtm$ is only affected for level positions around $\epsd = -(U + \Ud)/2$ in the SQD level interval $-2U < \epsSQD < 0$. We pointed out in Sec.~\ref{sec_singledot} that the fermion-parity mode is primarily excited for gate pulses shifting the dot either from $P_\mrm{d,0}^\mrm{in} > 0$ to the regime $P_\mrm{d,2}^\mrm{st} > 0$, where $m(t)$ and thus $\lamm$ primarily govern the zero-occupation probability $P_\mrm{d,0}(t)$, or from $P_\mrm{d,2}^\mrm{in} > 0$ to $P_\mrm{d,0}^\mrm{st} > 0$, with $m(t)$ describing the double-occupation probability $P_\mrm{d,2}(t)$ instead. So only by hampering the exponential decay of $P_\mrm{d,0}(t)$ for the first, and of $P_\mrm{d,2}(t)$ for the second kind of state transition can the SQD 
suppress the fermion-parity rate. However, as clarified by Fig.~\ref{fig_pseudospin}(c), gate pulses to induce such transitions, e.g., from $\epsdin > 0$ to a final dot level sufficiently far below $\epsd = -(U + \Ud)/2$, cause at least one electron to tunnel between dot and reservoir independently of the SQD state. More precisely, we add either one or two electrons to an empty dot in a regime where $m(t) = P_\mrm{d,0}(t)$, or remove a single or two electrons from a doubly occupied dot for $m(t) = P_\mrm{d,2}(t)$. Hence, neither the decay of $m(t) = P_\mrm{d,0}(t)$ for the first, nor the decay of $m(t) = P_\mrm{d,2}(t)$ for the second kind of process is affected by the SQD. So in agreement with Eq.~\eqref{eq_master_fermionparity}, only thermal fluctuations that are most likely to inhibit the relaxation for final dot levels close to $\epsd = -(U + \Ud)/2$ can cause a small deviation of the effective fermion-parity rate $\lamtm$ from the original one $\lamm$.\par

Yet, the mere existence of a weak suppression of $\lamtm$ for certain level positions is interesting, as the original rate $\lamm$ does not depend on any parameter apart from the tunnel coupling. It is shown in Ref. \onlinecite{Saptsov13} that the fermion-parity rate is rigorously protected against parameter changes in the absence of a detector. However, the capacitive coupling between dot and SQD accounted for here leads to a \textit{mutual} dependence of both subsystems on each other, and thus to memory effects that also slightly influence the fermion-parity rate: we do find a dependence, but the weakness of the effect indicates that some robustness remains.

\subsection{Measurement of the relaxation rates}\label{sec_measurement}

In this section, we demonstrate that it is possible to extract the original charge rate $\lamc$ and the dot fermion-parity rate $\lamm = 2\GamC/\hbar$ from SQD current measurements. We will emphasize, in particular, that measurements  of the bare rates are possible even in the presence of back-action effects.

\subsubsection{SQD tunnel-current}\label{sec_detector_current}
As it is our aim to infer the \textit{dot} dynamics from the time evolution of the SQD current, it is desirable to exclude the time-dependent influence of the SQD itself, namely the displacement current due to the relaxation of the SQD charge, from the measured signal. One way to suppress the influence of this displacement current on the measurement current is to actually perform \textit{two} measurements of the current in either the left lead $(I_\mrm{L})$ or right lead $(I_\mrm{R})$, and to take the second measurement with inverted source-drain voltage, $\mu_\mrm{L} \leftrightarrow \mu_\mrm{R}$.\footnote{In case of equal tunnel couplings, $\GamL = \GamR$, one can alternatively measure the symmetrized current $I(t) = (I_\mrm{L}(t) - I_\mrm{R}(t))/2$ in order to eliminate the impact of SQD displacement currents.} To see this explicitly, we label this second measurement as $I^\prime$ and calculate the difference between the first and second measurement using the current formula 
Eq.~\eqref{FormalismTunnelCurrentFinal},
\begin{eqnarray}
 I(t) & = & I_\mrm{L/R}^{\phantom{\prime}}(t) - I_\mrm{L/R}^\prime(t) \notag\\
 & = & \pm\frac{e\Gamma_\mrm{L/R}}{\hbar} \left\{\vphantom{\bbfR{+}}\left[\fL{+}-\fR{+}\right]P_\mrm{d,0}(t) +\left[\bbfL{+}-\bbfR{+}\right]P_\mrm{d,2}(t)\right.\notag\\
 && \phantom{\pm\frac{e\Gamma_\mrm{L/R}}{\hbar}}+\left.\left[\bfL{+}-\bfR{+}\right]\left(\vphantom{\bfL{+}}P_\mrm{d,\uparrow}(t) + P_\mrm{d,\downarrow}(t)\right)\vphantom{\left[\bbfL{+}-\bbfR{+}\right]}\right\}\label{eq_current_final}\ ,
\end{eqnarray}
where the positive/negative sign applies to lead L/R. The time dependence of $I(t)$ is hence solely determined by the dot dynamics, making it appropriate for the detection of the latter.\par

If the dot is empty at a certain time $t$, we have $P_\mrm{d,0}(t) = 1$ and thus $P_\mrm{d,\sigma}(t) = P_\mrm{d,2}(t) = 0$. Consequently, charges can already tunnel through the SQD if its level lies in the transport window $\mu_\mrm{R} < \epsSQD < \mu_\mrm{L}$. For a singly occupied dot, one finds  $P_\mrm{d,\uparrow}(t) +P_\mrm{d,\downarrow}(t) = 1$. Due to the Coulomb repulsion from the dot, the required SQD level energy for a current to flow through the SQD rises by the capacitive coupling strength $\UC$. 
Consequently, the relation $\mu_\mrm{R} < \epsSQD + \UC < \mu_\mrm{L}$ must be fulfilled to enable a sizable current. A completely filled dot with $P_\mrm{d,2}(t) = 1$ increases the necessary energy by $2\UC$, so that the current mostly flows for $\mu_\mrm{R} < \epsSQD + 2\UC < \mu_\mrm{L}$. This sensitivity of the current to the charge state of the quantum dot is the basis of the measurement process.\par

As $I(t)$ only depends on local dot probabilities, its time evolution after a gate step-pulse is not affected by the global fermion-parity rate $\lambda_{\text{gfp}} = (\Gamma + 2\GamC)/\hbar$, see Sec.~\ref{sec_global_fp}. The fact that $P_\mrm{d,\uparrow}(t)$ and $P_\mrm{d,\downarrow}(t)$ contribute equally to the current means that its decay is also not influenced by the dot spin $\sigma(t)$. Therefore the complete current dynamics is determined by the charge sector of the master equation, governed by $\mat{A}_\mrm{c}$ as discussed in Sec.~\ref{sec_charge}. More precisely, the time-dependent deviation of the current $I$ from its stationary limit $I^\mrm{st}$, namely, the current signal $\Delta I(t) = I(t) - I^\mrm{st}$, depends on the dot charge decay $\Delta n_\mrm{d}(t)$ and the dot fermion-parity quantity $\Delta m(t)$ according to
\begin{eqnarray}
 \Delta I(t) & = & \pm\frac{e\Gamma_\mrm{L/R}}{\hbar}\Big\{\left[(\delta\!f_\mrm{L}-\delta\!f_\mrm{R})-(\delta\!\bar{f}_\mrm{L}-\delta\!\bar{f}_\mrm{R})\right]\Delta m(t)\notag\\
  & + &\frac{(\delta\!{f}_\mrm{L} - \delta\!{f}_\mrm{R})\fd{-} + (\delta\!\bar{f}_\mrm{L} - \delta\!\bar{f}_\mrm{R})\fdUd{+}}{\fd{-} + \fdUd{+}} \Delta n_\mrm{d}(t)\Big\}.\notag\\ \label{eq_current_signal}
\end{eqnarray}
Since both $n_\mrm{d}(t)$ and $m(t)$ in general do not correspond to decay eigenmodes, $\Delta I(t)$ is, in principle, influenced by a combination of all four relaxation rates $\lamcm,\lamtm,\lamSET,\lamcp$. The contributions decaying at $\lamSET$ or $\lamcp$ are, however, irrelevant on the dot time scale, as we consider the measurement limit $\GamC/\Gamma \ll 1$. Moreover, for gate pulses which only change the dot level position $\epsdin \rightarrow \epsd$ but leave \textit{all} other parameters, in particular $\epsSQD$, fixed, the relative influences $a_\mrm{\widetilde{SQD}}$ and $a_+$ of the decay modes governed by $\lamSET$ and $\lamcp$ in the eigenmode expansion of $\Delta I(t)$ are smaller than $a_\mrm{\widetilde{c}}$ and $a_\mrm{\widetilde{fp}}$ associated with $\lamcm$ and $\lamtm$ by approximately two orders of magnitude. It is therefore justified to approximate
\begin{equation}\label{eq_current_decay}
 \Delta I(t) \approx a_\mrm{\widetilde{c}}e^{-\lamcm t} + a_\mrm{\widetilde{fp}}e^{-\lamtm t}\ .
\end{equation}
As we aim to individually extract the original charge rate $\lamc$ and the fermion-parity rate $\lamm$, the remaining task is to find parameter sets that fulfill either $a_\mrm{\widetilde{c}} \neq 0, a_\mrm{\widetilde{fp}} = 0$  and $\lamcm = \lamc$, or $a_\mrm{\widetilde{c}} = 0, a_\mrm{\widetilde{fp}} \neq 0$ and $ \lamtm = \lamm$.

\subsubsection{Measurement protocols}\label{sec_measurement_protocols}
The initial state of the combined d-SQD system is assumed to be given by the stationary state with respect to the system parameters before the switch, $\Pin = \Pst(\epsdin,\epsSQDin,\dotsc)$. The switch is realized by an instantaneous shift of the dot level position $\epsdin \rightarrow \epsd$; all other parameters are kept constant during the measurement.
To detect the dot dynamics induced by this gate pulse, a dc source-drain voltage $e\Vsd = U$ is applied to the leads, with $\mu_\mrm{L} = \mu_\mrm{C} = 0$ and $\mu_\mrm{R} = -e\Vsd$.\footnote{It is experimentally often advantageous to apply a finite frequency voltage in the rf-regime to the detector. However, if we assume SQD tunnel couplings $\Gamma/\hbar$ much larger than the typical voltage modulation frequency, the system still behaves as in the dc regime.} We set the temperature $\kBT = \Ud/10$ sufficiently low and the capacitive coupling $U = 5\Ud/8$ high enough for the SQD to be able to sense the individual dot occupation probabilities $P_\mrm{d,0}(t),P_\mrm{d,\uparrow}(t)+P_\mrm{d,\downarrow}(t),P_\mrm{d,2}(t)$. Finally, we set $\Gamma_\mathrm{C}=\Gamma/10$ to approximate the measurement limit. Since $\Gamma/\kBT \ll 1$ and thus $U/\Gamma \gg 1$, we describe a \textit{strong} measurement.

In order to measure either the charge rate $\lamc$ or the fermion-parity rate $\lamm$, it is crucial that we can distinguish between both rates. As the rates are quasi\footnote{Strictly speaking, we always find $\lamcm < \lamtm$, even if the difference is small} degenerate in the regime of single-occupation, each decay process discussed in the following must be induced by a gate pulse into either the zero occupation or the double occupation regime, namely $\epsd/\kBT \gg 1$ or $-(\epsd+\Ud+U)/\kBT \gg 1$. For these level positions, the rates $\lamc \approx \GamC/\hbar$ and $ \lamm = 2\GamC/\hbar$ differ strongly.  But even more importantly, the corresponding decay modes are not affected by back-action and in particular do not couple, as evident from Sec.~\ref{sec_charge}. 
The task of extracting the individual rates therefore becomes equivalent to suppressing the contribution of either the charge mode $\propto\Delta n_\mrm{d}(t)$ or the fermion-parity mode $\propto\Delta m(t)$ in the current signal decay according to Eq.~\eqref{eq_current_signal}. 

\begin{figure}
 \includegraphics[width=\linewidth]{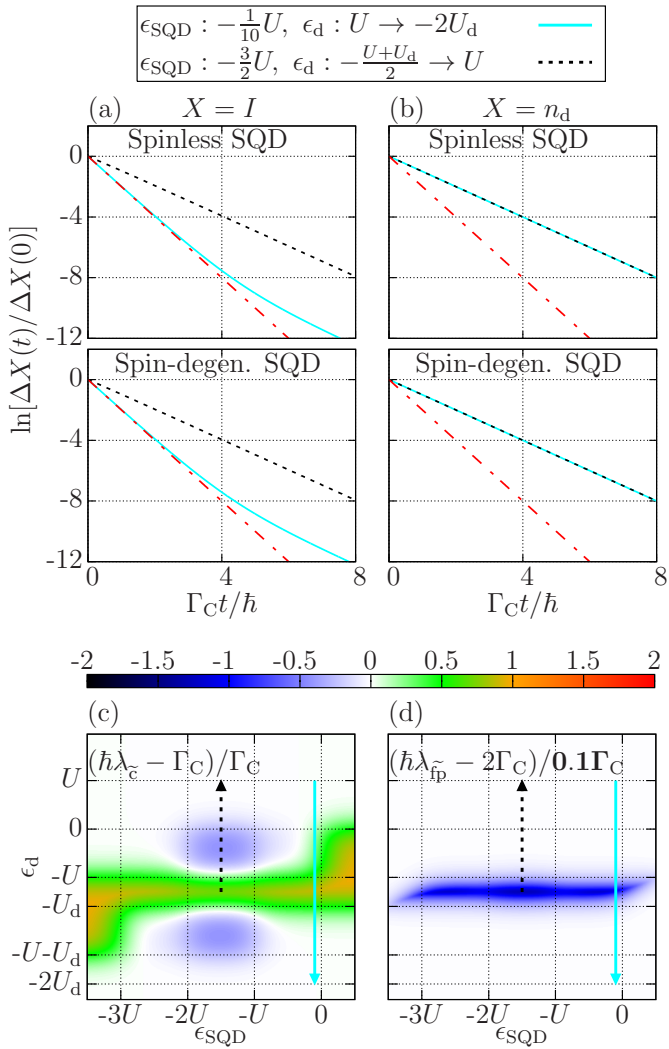}
 \caption{Logarithm of (a) the current signal $\Delta I(t)$ and (b) the dot charge deviation $\Delta n_\mrm{d}(t)$, as a function of time for two different gate pulses, both for a spinless and spin-degenerate SQD. The quantities are normalized by their respective initial values $\Delta I(0)$, $\Delta n_\mrm{d}(0)$. The red dash-dotted line represents in each case the decay of $\Delta m(t)/\Delta m(0)$. (c,d) The two gate pulses are visualized as arrows in a plot of the effective charge- and dot fermion-parity rate $\lamcm$, $\lamtm$ as a function of $\epsSQD,\epsd$. The starting/end point of each arrow indicates the level positions before/after the switch, with the initial state $\Pin = \Pst(\epsSQDin,\epsdin,\dotsc)$. The system parameters are: $\mu_\mrm{L} = \mu_\mrm{C} = 0$, $\mu_\mrm{R} = -e\Vsd = -U$, $\GamC = \Gamma/10$, $\Ud = 8U/5$, $\kBT = \Ud/10 = 0.16U$. The SQD on-site interaction strength is $U_\mrm{SQD} = 2U$. As before, we must require $\Gamma/\kBT \ll 0.1$, even though not explicit in the 
calculation.}
 \label{fig_measurement} 
\end{figure}

In the following, we address the individual detection of $\lamc \rightarrow \GamC/\hbar$ and $\lamm = 2\GamC/\hbar$ by motivating and subsequently discussing a concrete example of a gate pulse and a measurement scheme for each rate. Each pulse is visualized by an arrow in Figs.~\ref{fig_measurement}(c),(d), which show the effective decay rates $\lamcm$ and $\lamtm$ as a function of the level positions for $e\Vsd = U$. Both arrows start at the initial level positions ($\epsdin,\epsSQDin$) and point to the final level positions ($\epsd,\epsSQD$) of the corresponding gate pulse. The graphs of Fig.~\ref{fig_measurement}(a),(b) display how the current signal $\Delta I(t)$ and the dot charge $\Delta n_\mrm{d}(t)$ evolve as function of time for each gate pulse. We plot $\Delta I(t)/\Delta I(0)$ and $\Delta n_\mrm{d}(t)/\Delta n_\mrm{d}(0)$ logarithmically, as the negative slope of the curves equals the relaxation rate at which the corresponding quantity decays. To demonstrate that an additional SQD spin degree of 
freedom does not substantially influence the measurement, we also show numerical results using a spin-degenerate SDQ, see details in Appendix \ref{app_SDSQD}.

1. \textit{Measurement of the dot fermion-parity rate}:  
Inducing an exponential decay of $\Delta m(t)$ in order to measure the fermion-parity rate $\lamm = 2\GamC/\hbar$ is most effectively accomplished by shifting the dot either from an empty to a doubly occupied state or vice versa, see Sec.~\ref{sec_singledot}. Since, however, any dot-state transition automatically triggers the charge $\Delta n_\mrm{d}(t)$ to decay as well, we have to find parameters that minimize the sensitivity of the SQD to $\Delta n_\mrm{d}(t)$, that is the prefactor in front of $\Delta n_\mrm{d}(t)$ in the current signal formula, Eq.~\eqref{eq_current_signal}.

In the example considered here, we gate-pulse the dot from zero to double occupation, namely by a shift from $\epsd =U\rightarrow \epsd = -2 \Ud$ as illustrated by the blue arrows in Fig.~\ref{fig_measurement}(c),(d). For this process, we have $m(t) = P_\mrm{d,0}(t)$. We are therefore required to set the SQD sensitive to \textit{only} the zero occupation probability $P_\mrm{d,0}(t)$ in order to exclude any influences from the charge decay. This is here achieved by setting $\epsSQD = -U/10 $, which is very close to, but still below the chemical potential of the left lead.\par

The blue solid curves in Fig.~\ref{fig_measurement}(a),(b) show the resulting time dependence of the current signal (a) and the dot charge (b). For times $\GamC t/\hbar \lesssim 3$, $\Delta I(t)$ decays at a rate given by almost $2\GamC/\hbar$, whereas the charge decays at the unperturbed charge rate $\lamc = \GamC/\hbar$. This clearly indicates that we indeed measure the fermion-parity rate and not the charge rate. However, for later times, $\GamC t/\hbar > 3$, the relaxation rate of the current signal gradually decreases to $\GamC/\hbar$. The reason is that we cannot eliminate the influence of the charge decay mode entirely; as this decay process dies out at a slower rate $\lamc = \GamC/\hbar < \lamm$, it must eventually dominate the time-dependent decay of the current signal. It is however only important to keep the charge rate influence negligible up to times where $\Delta I(t)$ has in practice already decayed (which is essentially the case at $\GamC t/\hbar \approx 3$). Improvements are possible by 
increasing the ratio $U/\kBT$ determining the resolution of the 
detector. \par

2. \textit{Measurement of the charge relaxation rate}: One can tell from the prefactor in front of $\Delta m(t)$ in the current signal formula \eqref{eq_current_signal} that, as long as the SQD current depends on the dot state, it is always sensitive to the fermion-parity decay mode. Hence the only way to measure the pure unperturbed charge rate is to consider a gate pulse that does not cause $m(t)$ to decay at all, namely $\Delta m(t) = 0$. This means that we must start with a singly occupied dot and apply a gate pulse that drives the dot into zero or double occupation. 
Here, we shift the dot into the empty regime as illustrated by the black (dotted) arrows in Fig.~\ref{fig_measurement}(c),(d) ranging from $\epsd =-(U+\Ud)/2\rightarrow \epsd = U$. To measure a current decay for this gate pulse, the SQD must be sensitive to either $P_\mrm{d,0}(t)$ or $P_\mrm{d,\uparrow}(t)+P_\mrm{d,\downarrow}(t)$. We choose the latter in our example, and set the SQD level position $\epsSQD = -3U/2$ in the middle of the transport window $\mu_\mrm{R} = -e\Vsd < \epsSQD + U < \mu_\mrm{L} = 0$ accordingly.\par

The black dotted curve for $\Delta I(t)$ in Fig.~\ref{fig_measurement}(a) coincides almost perfectly with the curve for $\Delta n_\mrm{d}(t)$ in (b), and consequently enables us to extract the unperturbed charge relaxation rate $\lamc = \GamC/\hbar$. Since the chosen gate pulse ideally does not excite the fermion-parity mode, the resolution of the SQD, namely the ratio $U/\kBT$, does not have to be as high as for the measurement of $\lamm$. In fact, too large ratios of $U/\Ud$ make it even more difficult to ensure that the dot is singly occupied in the initial state, and consequently, to avoid exciting the fermion-parity decay mode.

\section{Conclusions}\label{conclusions}

We have studied the read-out of the relaxation rates of a quantum dot, driven out of equilibrium by a fast parameter change, via the current through a capacitively coupled sensor quantum dot (SQD). We therefore studied in detail the dynamics of the \textit{full} dot-SQD system, based on a generalized master equation approach, revealing all seven relaxation rates of the combined system. Subsequently, we extracted the sensitivity of the measurement current on the dot state and the back-action effects of the SQD on the relaxation behavior of charge, spin, and a fermion-parity-related quantity, $m$, of the dot. We therefore explored the relaxation behavior in the full parameter space of quantum dot and SQD.

Our main interest was devoted to the possibility of detecting the charge relaxation rate of the quantum dot, $\lambda_\mathrm{c}$ on one hand, and the up to now experimentally unrevealed relaxation rate, $\lambda_\mathrm{fp}$, which governs the
dynamics of the fermion-parity related quantity $m$, on the other hand. 

We could show that back-action effects lead to capacitive shifts in the gate dependence of the dot relaxation rates, to a Coulomb-blockade induced suppression of relaxation rates in certain parameter regimes and, in general, to a mixing of the dynamics of the charge and the fermion-parity related quantity $m$. However, the fermion parity rate, $\lambda_\mathrm{fp}$, remains well protected even in the presence of the sensor quantum dot.

Based on these observations we could devise measurement protocols making use of the sensitivity of the SQD on the charge states of the quantum dot. In particular, the charge relaxation rate could be shown to be detected when switches of the dot between dot states differing by one electron charge are performed. Most interestingly, also the read-out of the fermion-parity relaxation rate was shown to be possible, when detecting the relaxation after switches between dot states differing by two electron charges.

\acknowledgments
We would like to thank Michael Hell, J\"urgen K\"onig, Mikhail Pletyukhov, Roman Saptsov and Maarten Wegewijs for interesting discussions and helpful comments on the manuscript. We acknowledge financial support from the Ministry of Innovation NRW (J.S., J.S., L.D. C.-P.), the Swedish Research Council (J.S., J.S.), and the European Commission (STREP PICC) and the Alexander von Humboldt Foundation (L.D. C.-P.). M.G. acknowledges the hospitality at RWTH Aachen during a visit financed through the ERS international programme.

\section{Appendix}
\appendix

\section{Spin eigenvectors}\label{app_spinvec}

In Sec.~\ref{sec_spinsector}, we have analytically determined the spin relaxation dynamics  by diagonalizing the spin master equation Eq.~\eqref{eq_spin_master}. Here, we want to state the left and right eigenvectors. Defining $\gamma_\mrm{C} = \GamC/\Gamma$ and
\begin{equation}
g_\sigma = \sqrt{\scriptstyle\left[1 + \gamma_\mrm{C}\left(\vphantom{\bfL{+}}\Dfd - \DfdUd\right)\right]^2 - 4\gamma_\mrm{C}\bfS{+}\left[\Dfd - \DfdUd\right]}\ , 
\end{equation}
we can write the left and right eigenvectors belonging to the rates $\lamsigpm$ as follows:
\begin{eqnarray*}
&& \hat{l}^+_\sigma = \\
&&    \frac{\bfS{-} - \bfS{+} + \gamma_\mrm{C}\left[\Dfd - \DfdUd\right] - g_{\sigma}}{2g_{\sigma}}\hat{l}^0_\sigma - \frac{\bfS{-}}{g_{\sigma}}\hat{l}^1_\sigma \\
&&\hat{l}^-_\sigma  =   - \frac{\bfS{-} - \bfS{+} + \gamma_\mrm{C}\left[\Dfd - \DfdUd\right] - g_{\sigma}}{2\bfS{+}}\hat{l}^{1}_\sigma+ \hat{l}^0_\sigma\\
&&\mbf{P}^+_{\sigma}   =   - \frac{\bfS{-} - \bfS{+} + \gamma_\mrm{C}\left[\Dfd - \DfdUd\right] + g_{\sigma}}{2\bfS{-}}\mbf{P}_\sigma^1  + \mbf{P}_\sigma^0\\
&&\mbf{P}^-_{\sigma} =  \\
 &&\frac{\bfS{-} - \bfS{+} + \gamma_\mrm{C}\left[\Dfd - \DfdUd\right] + g_{\sigma}}{2g_{\sigma}}\mbf{P}_\sigma^0  + \frac{\bfS{+}}{g_{\sigma}}\mbf{P}_\sigma^1\ ,
\end{eqnarray*}
with the row- and column vectors
 \begin{eqnarray*}
  \hat{l}^0_\sigma & = & (0,1,-1,0,0,0,0,0)\\
  \hat{l}^1_\sigma & = & (0,0,0,0,0,1,-1,0)\\
  \mbf{P}^0_{\sigma} & = & (0,1/2,-1/2,0,0,0,0,0)^T\\
  \mbf{P}^1_{\sigma} & = & (0,0,0,0,0,1/2,-1/2,0)^T
 \end{eqnarray*}
expressed in the eigenbasis of the d-SQD system. 
\section{Measurements with spin-degenerate SQD}\label{app_SDSQD}

In Sec.~\ref{sec_hamiltonian}, we claimed that for sufficiently large SQD on-site interactions, the outcome of a measurement of the dot dynamics with a spin-degenerate single-level SQD does not deviate significantly from the results using a spinless SQD. While a rough comparison of the two cases shown in Figs.~\ref{fig_measurement}(a),(b) already supports our statement, we here want to give a more quantitative analysis for the gate pulses considered in Sec.~\ref{sec_measurement_protocols}. More precisely, we compare the slopes of the time-dependent current and charge signal curves
\begin{equation}
 s_X(t) = \ddt{}\ln\left|\frac{\Delta X(t)}{\Delta X(0)}\right|,\ X = I,n_\mrm{d},\ 
\end{equation}
and thus the effective relaxation rate governing the exponential decay for both a spinless (SL) as well as a spin-degenerate (SD) SQD. 

The relative difference $(s_{X,\mrm{SD}}(t) - s_{X,\mrm{SL}}(t))/|s_{X,\mrm{SL}}(t)|$ between the time-dependent effective relaxation rates of the spinless and spin-degenerate system, both for the current signal $\Delta I(t)$ and the dot charge $\Delta n_\mrm{d}(t)$ are shown in Fig.~\ref{fig_comparison} for the two relaxation processes that reveal the bare charge rate as well as the fermion-parity rate (see also Fig.~\ref{fig_measurement}). The system parameters are equal to those given in Fig.~\ref{fig_measurement}. In particular, the SQD on-site repulsion strength $U_\mrm{SQD} = 2U > \Ud$ is defined to be large enough to prevent the SQD from being doubly occupied; any $U_\mrm{SQD}$ larger than the chosen one will thus lead to the same results.\par

\begin{figure}
 \includegraphics[width=\linewidth]{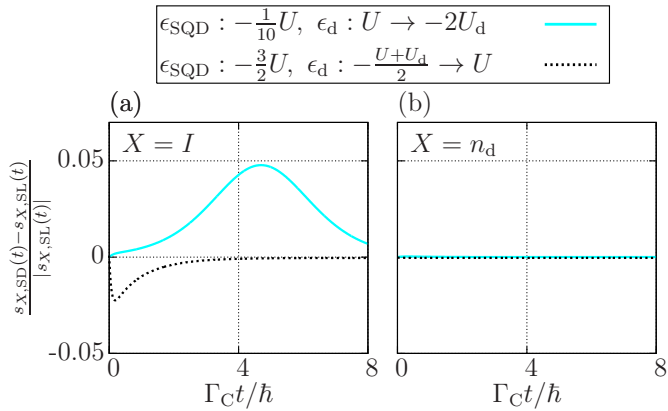}
 \caption{Relative difference between the slopes $s$ of the logarithmic current signal and dot charge curves as obtained for a spinless SQD, see Fig.~\ref{fig_measurement}, and a spin-degenerate SQD. System parameters are: $\mu_\mrm{L} = \mu_\mrm{C} = 0$, $\mu_\mrm{R} = -e\Vsd = -U$, $\GamC = \Gamma/10$, $\Ud = 8U/5$, $U_\mrm{SQD} = 2U$, $\kBT = \Ud/10 = 0.16U$. We assume $\Gamma/\kBT \ll 0.1$.}
 \label{fig_comparison}
\end{figure}

For the shift from the single- to the zero-occupation regime (black dotted line), intended to expose the relaxation rate $\GamC/\hbar$, the relative deviation between both results for the current signal is $\leq2.5\%$ for all times $t$. The dot charge decay is not affected at all by the additional spin degree of freedom in the SQD. The difference between a spinless and spin-degenerate detector is therefore negligible. For the process used to measure the fermion-parity rate, displayed by the blue solid curve, the deviation of the SD current signal slope from the SL result gradually increases to almost $5\%$ for $t = 5\hbar/\GamC$, namely when the influence of the charge rate becomes relevant. However, we have pointed out that the current signal has pratically already decayed after such a time span. For the relevant times between $\hbar/k_\mathrm{B} T$ and $3\hbar/\GamC$, the relative difference is still below $2.5\%$. The charge decay is again not significantly affected by the SQD spin.\par

Moreover, let us point out that whereas the observed difference in the black dotted curve is most significant for small times on the order of the inverse SQD tunneling rate $\hbar/\Gamma$, the blue curve deviates most strongly for larger times on the dot time scale. This means that the deviation in the detection of the charge rate is at least partly caused by the way the relaxation modes closely related to the SQD affect the current decay, whereas the difference observed for the fermion-parity rate must be an effect of the Coulomb interaction inside the SQD, leading e.g. to energy-shifted conductance resonances. 

In summary, it is justified to simplify the description of the SQD by assuming it to be spinless, at least on time scales that are relevant for the measurement.

%

\end{document}